# Tunneling magneto thermo power in magnetic tunnel junction nanopillars

*N. Liebing[1], S. Serrano-Guisan[1*], K. Rott[2], G. Reiss[2], J. Langer[3], B. Ocker[3] and H. W. Schumacher[1]*

1) Physikalisch-Technische Bundesanstalt, Bundesallee 100, D-38116 Braunschweig, Germany.
2) University of Bielefeld Department of Physics Universitätsstr. 25 33615 Bielefeld Germany.
3) Singulus AG, Hanauer Landstrasse 103, D-63796 Kahl am Main, Germany

* Corresponding author:
E-mail:    santiago.serrano-guisan@ptb.de
Phone:    +49 (0)531 592 2439,    fax:    +49 (0)531 592 69 2439

## Abstract

We study the tunneling magneto thermo power (TMTP) in CoFeB/MgO/CoFeB magnetic tunnel junction nanopillars. Thermal gradients across the junctions are generated by a micropatterned electric heater line. Thermo power voltages up to a few tens of µV between the top and bottom contact of the nanopillars are measured which scale linearly with the applied heating power and hence with the applied temperature gradient. The thermo power signal varies by up to 10 µV upon reversal of the relative magnetic configuration of the two CoFeB layers from parallel to antiparallel. This signal change corresponds to a large spin-dependent Seebeck coefficient of the order of 100 µV/K and a large TMTP change of the tunnel junction of up to 90 %.

**Article**

Transport coefficients in ferromagnetic materials are spin-dependent.[1] The giant magneto resistance[2] and the tunneling magneto resistance (TMR)[3] are the most prominent examples of spin dependent *electronic* transport. Their discovery boosted *spintronics*[4] with important applications e.g. in data storage. While spintronics relies on spin dependent *charge* transport also spin dependent *heat* transport can become important when, e.g., a high current density in magnetic nanodevices creates a significant temperature gradient. The combination and coupling of heat, charge, and spin currents in magnetic nanostructures has recently opened a highly active line of research now referred to as *spin caloritronics*[5] with important discoveries such as the spin-Seebeck effect[6,7,8] or thermally driven spin injection[9]. Furthermore, thermal spin transfer torque was predicted to enable highly efficient magnetization reversal in magnetic nanodevices by thermal gradients[10,11,12] and first experimental evidence has recently been provided.[13]

For *spintronics* CoFeB/MgO/CoFeB magnetic tunnel junctions (MTJ) are one of the most attractive systems as they can be reliably produced by sputter deposition[14] and show very high TMR ratios. The latter results from the half-metallic character of the coherently tunneling $\Delta_1$ states in the ferromagnetic electrodes.[15] With respect to *spin caloritronics* applications recent *ab-initio* studies also predicted very high spin-dependent Seebeck coefficients of 150 µV/K and correspondingly high tunneling magneto thermo power (TMTP) ratios of the MTJ.[16] Furthermore magnon-assisted tunneling might contribute to TMTP especially in nanosized, mesoscopic MTJs.[17,18] An experimental confirmation of these predictions would strongly impact materials research in the field of *spin caloritronics*.

Here, we experimentally study TMTP in CoFeB/MgO/CoFeB nanopillars in the presence of thermal gradients across the MTJ. We find a strong increase of the thermo power voltage upon reversal of the magnetic configuration of the two CoFeB layers from parallel to antiparallel. The measured increase of up to 10 µV corresponds to large spin-dependent Seebeck coefficients of the order of 100 µV/K and to large TMTP ratios of the MTJ of 90 % making this material system a highly attractive candidate for future *spin caloritronic* applications.

The MTJ stacks are sputter deposited in a Singulus NDT Timaris cluster tool on a Si wafer capped with 100 nm $SiO_2$. The MTJ stack consists of a bottom contact (BC) of 3nm Ta/90nm Cu/5nm Ta, a pinned layer (PL) stack comprising a synthetic antiferromagnet of 20nm PtMn/2nm $Co_{60}Fe_{20}B_{20}$/0.75nm Ru/2nm $Co_{60}Fe_{20}B_{20}$, the 1.5 nm MgO tunnel junction, the free layer (FL) of 3nm $Co_{60}Fe_{20}B_{20}$, and a top contact of 10nm Ta/30nm Cu /8nm Ru. The completed stack is annealed for 90 minutes at 360° C in 1 T field and lithographically patterned into 160 × 320 nm² wide elliptic nanopillars. An electron micrograph of a typical device is shown in Fig. 1(b). Electrical top contacts (TC) to the nanopillar are provided by electron beam lithography and lift-off. To apply thermal gradients a 5 µm wide and 70 nm thick Au heater line (HL, marked by full lines) is patterned on top of the nanopillar and is separated from the TC by a 160 nm thick $Ta_2O_5$ dielectric. The 10 µm wide BC line (marked by dashed lines) is running underneath the HL. It acts as a heat sink to establish a well defined temperature gradient across the MTJ. The nanopillars typically show single domain magnetization reversal with uniaxial anisotropies between 8 mT $\leq \mu_0 H_k \leq 15$ mT, TMR ratios between 70 and 140 %, and resistance area products of the order of ~ 17 $\Omega\mu m^2$.[19]

We characterize the spin caloric properties of our MTJ nanopillars by magneto thermo electrical measurements. Such measurements have been previously used to characterize e.g. magnetic multilayers,[20] nanowires,[21] and granular systems.[22,23] The experimental setup is sketched in Fig. 1(a). Static fields up to $\mu_0 H_S$ = 60 mT are applied in arbitrary in plane orientations. Thermal gradients across the MTJ are generated by applying AC or DC heater currents up to $I_{heat}$ = 60 mA through the HL while $V_{TP}$ between BC and TC of one MTJ is measured. In our measurements a positive $V_{TP}$ corresponds to a positive voltage at the TC. For DC heater currents $V_{TP}$ is measured by a nanovoltmeter. For AC measurements at AC heater frequencies of 10 … 80 Hz $V_{TP}$ is measured by a lock-in detection at the second harmonic. Note that $I_{heat}$ also generates an easy axis field $H_{heat}$. For DC measurements $H_{heat}$ is compensated by an external easy axis field $H_X$ whereas for AC measurements such compensation is not possible in our present setup. Hence while AC experiments allow determining $V_{TP}$ with better signal-to-noise ratio the measured $V_{TP}(H_S)$ will always be averaged over an oscillating easy axis field $\pm H_{heat}$ of up to $\pm 6$ mT.

Fig. 2(a) shows an easy axis TMR loop (black) of one of the MTJ nanopillars (MTJ-3 in Table 1). The TMR is measured at a current bias of 100 μA in easy axis fields up to $\mu_0 H_X = \pm 25$ mT. The TMR change of 110 % occurs at the reversal from parallel (P) to antiparallel (AP) orientation of the magnetization of the 3 nm thick CoFeB free layer (FL) with respect to the pinned layer (PL) magnetization. In our experiments the PL magnetization is always oriented along the positive x-direction. Note that the sharp jumps of the TMR at the coercive fields $H_C^-$ and $H_C^+$ speak for a single domain reversal behavior of the MTJ FL which is confirmed by well defined switching asteroids in combined easy axis and hard axis fields (not shown; cp. Ref. 19, Fig. 1b for a typical result).

The red curve in Fig. 2(a) is a TMR loop taken under application of $I_{heat} = 38$ mA through the HL. Here, the effect of $H_{heat}$ is visible and $H_C^-$ and $H_C^+$ are both offset by about 3.5 mT to positive $H_X$. Measurements of the coercive field offsets $\Delta H_C^-$, $\Delta H_C^+$ up to heater currents of $I_{heat} = 60$ mA show a linear scaling of $H_{heat}$ with $I_{heat}$ with a slope of about 0.1 mT/mA (not shown). Note that no significant reduction of $H_C$ with $I_{heat}$ during heating is found in the applied current range of our experiments and thermally activated reversal can be neglected.

Fig. 2(b) shows three typical DC measurements of $V_{TP}$ of the same device. $V_{TP}$ is displayed for heater powers $P_{heat}$ of 21, 38 and 58 mW as function of the easy axis field $H_X$.

$H_{heat}$ is compensated by a static field offset. For all three curves $V_{TP}(P)$ in the P state is lower than $V_{TP}(AP)$ in the AP state with a maximum difference of $\Delta V_{TP} \approx 11$ µV. This yields an average TMTP ratio of the given device of TMTP = $\Delta V_{TP}/ V_{TP}(P) \approx 32\%$.

Fig. 3(a) shows the typical dependence of $V_{TP}(P,AP)$ on $P_{heat}$. DC and AC data agree well. Measurements were performed at $\mu_0 H_X = \pm 30$ mT in well defined P and AP configurations. $V_{TP}(P)$, $V_{TP}(AP)$ both scale linearly with $P_{heat}$ and hence with the temperature gradient $\Delta T_{MTJ}$ across the MTJ as expected. In Fig. 3(b) the TMTP ratio derived from the same data is plotted vs. $P_{heat}$. The AC data yields a constant TMTP of about 32% (full squares). Here, AC measurements have only been carried out up to $P_{heat}$ = 34 mW. Note that the TMTP derived from AC measurements is typically constant over the whole power range displayed in the figure. Also the TMTP derived from DC measurements (open squares) is constant for $P_{heat} \geq$ 20 mW with a comparable TMTP ≈ 29 %. The lower TMTP for $P_{heat} \leq$ 20 mW in the DC data can be attributed to an artifact resulting from uncompensated voltage offsets of the nanovoltmeter which are becoming significant at low $P_{heat}$.

Comparison of our data to the predicted high spin-dependent Seebeck coefficients requires an estimate of the temperature gradient $\Delta T_{MTJ}$ over the MgO MTJ. Therefore the HL resistance $R_{HL}$ is measured as function of $P_{heat}$ (Fig. 3(c)). From the increase of $R_{HL}$ with $P_{heat}$ the increase of the HL temperature $\Delta T_{HL}(P_{heat})$ can be derived by $R_{HL} = R_{HL,0} (1+\alpha \cdot \Delta T_{HL})$, where $R_{HL,0}$ is the resistance at room temperature ($P_{heat}$ = 0 mW) and $\alpha = 3.9 \cdot 10^{-3}$ K$^{-1}$ is the temperature coefficient of the HL material Au.[24] $\Delta T_{HL}(P_{heat})$ is plotted as red triangles in Fig. 3(c). For maximum $P_{heat}$ = 60 mW a maximum $\Delta T_{HL} \approx 15.2$ K is found. Note that in parallel no change of resistance of the BC line and hence of the BC temperature was found. $\Delta T_{HL}$ thus represents a good estimate of the temperature drop between HL and BC. Based on this the temperature distribution over the nanopillar structure was computed using a commercial finite element solver.[25] We use a two dimensional model of the pillar structure including contacts and insulating layers with thermal material parameters based on bulk literature values.[24,26,27] The simulations show that the dominant temperature drop occurs across the 160 nm dielectric between HL and TC. In Fig. 1(c) the temperature increase in the center of the pillar above the BC is plotted as function of layer thickness $t$. For the given maximum $P_{heat}$ = 60 mW one finds a temperature drop of $\Delta T_{MTJ} \approx 41$ mK across the MgO barrier (arrow). Based on this one can estimate a spin-dependent Seebeck coefficient of the MTJ of $S_{MTJ} = \Delta V_{TP}/\Delta T_{MTJ} \approx 250$ μV/K comparable to the predicted value of 150 μV/K.[16] Note, however, that due to the large uncertainty of $\Delta T_{MTJ}$ (e.g. due to the use of bulk values for the thermal conductivity of thin films) only an order of magnitude estimate of $S_{MTJ}$ is feasible.

The above *ab-initio* studies of TMTP in MgO based MTJs have also considered the angular dependence of $V_{TP}$.[16] Fig. 2(c,d) shows the measured angular dependence of (c) TMR and (d) $V_{TP}$ for a 360° in-plane rotation of the FL magnetization induced by a rotating static field $H_S$. $V_{TP}$ is measured with AC heater current. $V_{TP}$ well follows the typical $\cos(\phi-\pi)$ dependence of the TMR.[28] In the TMR two jumps are found around $\phi \approx \pm 90°$ when the free layer magnetization overcomes the hard axis. In $V_{TP}$ these jumps are smoothed out by the uncompensated AC heater field of ± 6 mT. Note that the measured angular dependence does not well follow the theoretical prediction of an almost constant $V_{TP}$ for $|\phi| < \pm 120°$ and a sharp increase near the AP orientation.[16] In contrast both in collinear and tilted field configurations of FL and PL $V_{TP}$ basically follows the field dependence of the TMR.

As listed in Table 1 our samples reveal TMTP ratios between 17 % and 42 % and TMR ratios between 79 % and 140 %. In the inset to Fig. 3(b) the TMTP ratios of the different samples are plotted as function of the TMR. No significant correlation of the amplitude of TMR and TMTP is found. The TMTP rather seems to scatter around a value of about 31%. Theoretically no close correlation of TMR and TMTP ratio is expected. While the TMR is sensitive to the density of states (DOS) of the two spin channels at the Fermi level the TMTP is sensitive to the *asymmetry* of the DOS.[16] Note that $V_{TP}$ and hence the above TMTP ratios also contain contributions of all non-magnetic layers of the devices. To determine these the MgO barriers of some devices were shortened by pinholes[29] by application of current stress resulting in a field independent resistance of ~ 50 Ω. Fig. 2(b) also shows $V_{TP,short}$ of the shortened MTJ-3 for $P_{heat}$ = 58 mW (dashed line). $V_{TP,short}$ is independent of field thereby unambiguously confirming the origin of TMTP at the MgO MTJ. Subtracting this background from $V_{TP}(P)$ yields a better estimate of the TMTP contribution of the CoFeB/MgO/CoFeB MTJ. The resulting values of $TMTP_{MTJ} = \Delta V_{TP}/(V_{TP}(P) - V_{TP,short})$ of e.g. 72 % (MTJ-2) and 90% (MTJ-3) are significantly higher than listed in Table 1.

Concluding we have experimentally demonstrated a large TMTP of up to 90% in CoFeB/MgO/CoFeB MTJs making this material system a promising candidate for future *spin caloritronics* applications. The derived spin-dependent Seebeck coefficients are comparable to *ab-initio* predictions while deviations from the predicted angular dependence are found.[16] For all devices $V_{TP}(AP)$ was larger than $V_{TP}(P)$. This agrees both with *ab-inito* theory and with predictions of TMTP due to magnon-assisted tunneling in half metallic *mesoscopic* MTJs which could be relevant for our MTJ nanopillars devices.[17,18] Here, future temperature dependent experiments could shine light on the origin of the TMTP in our devices as a strong temperature dependence and even a sign change of the spin-dependent Seebeck coefficient has been predicted by *ab-initio* theory.[16]

Note added: In a recent preprint similar results have been obtained in an experiment using optical heating of MgO based MTJs of several μm lateral dimensions.[30] We like to thank M. Münzenberg and Ch. Heiliger for stimulating discussions. We acknowledge funding by the EU IMERA-Plus Grant No. 217257 and by the DFG Priority Program SpinCaT.

**TABLE:**

| Name | $\mu_0 H_k$ (mT) | TMR (%) | TMTP (%) |
|---|---|---|---|
| **MTJ-1** | 6 | 79 | 32 |
| **MTJ-2** | 8 | 88 | 17 |
| **MTJ-3** | 15 | 110 | 32 |
| **MTJ-4** | 6.5 | 134 | 41 |
| **MTJ-5** | 8 | 137 | 30 |

**Table 1**

**N. Liebing et al.**

FIGURES :

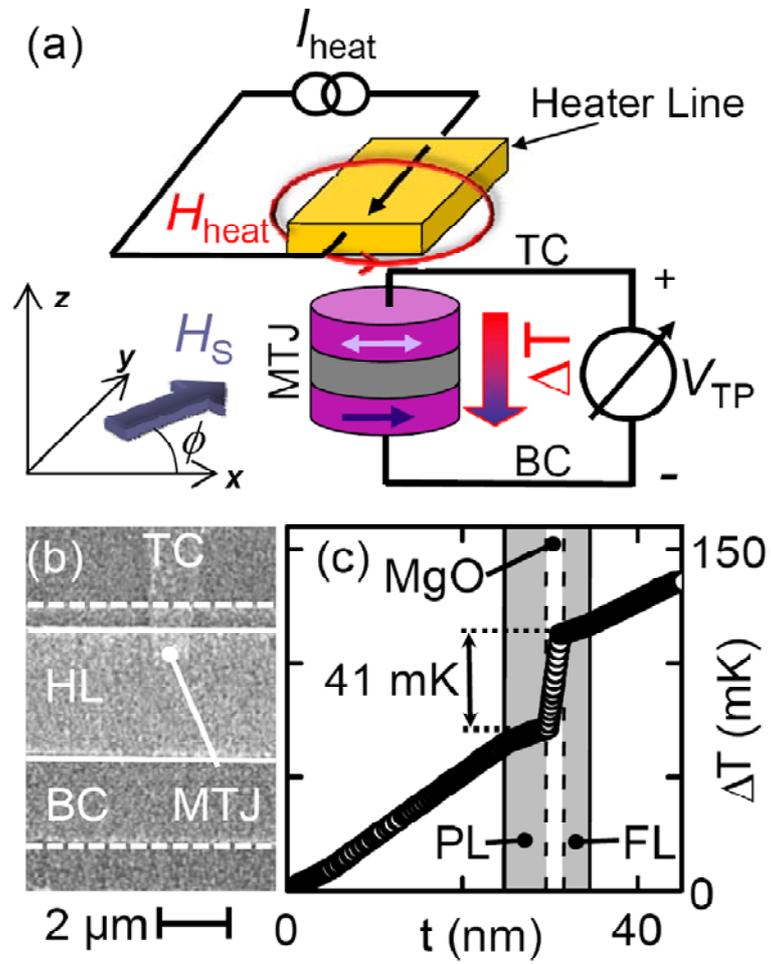

**Figure 1**
**N. Liebing et al.**

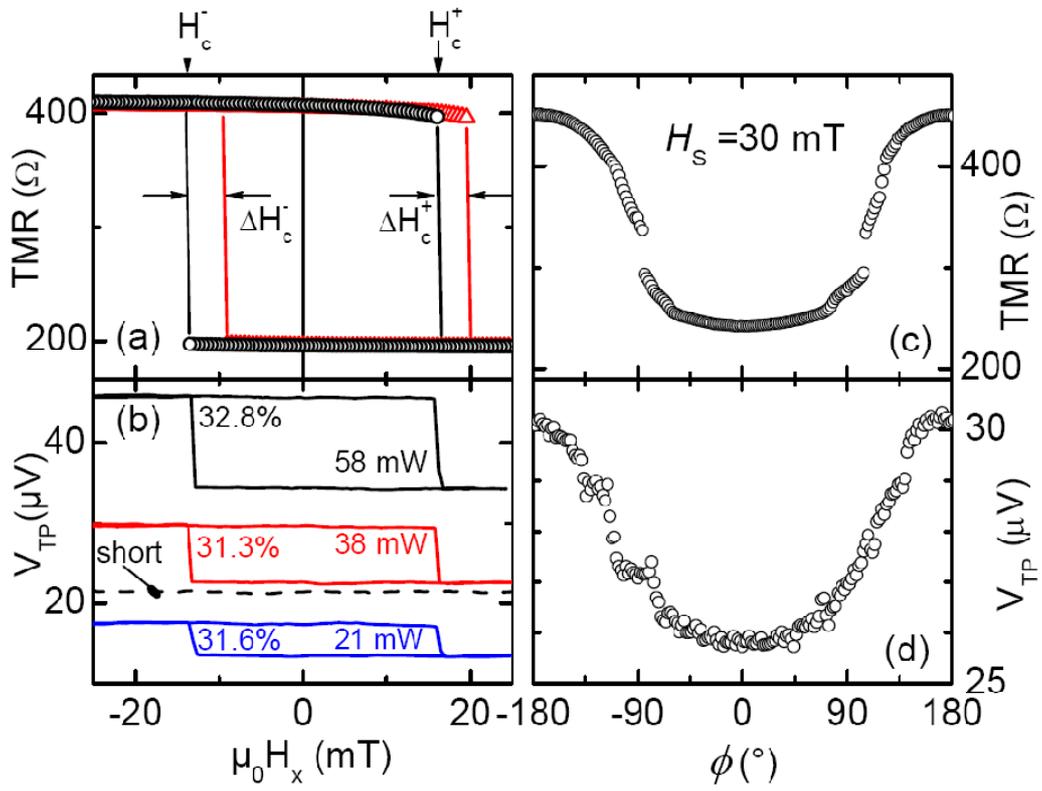

**Figure 2**

**N. Liebing et al.**

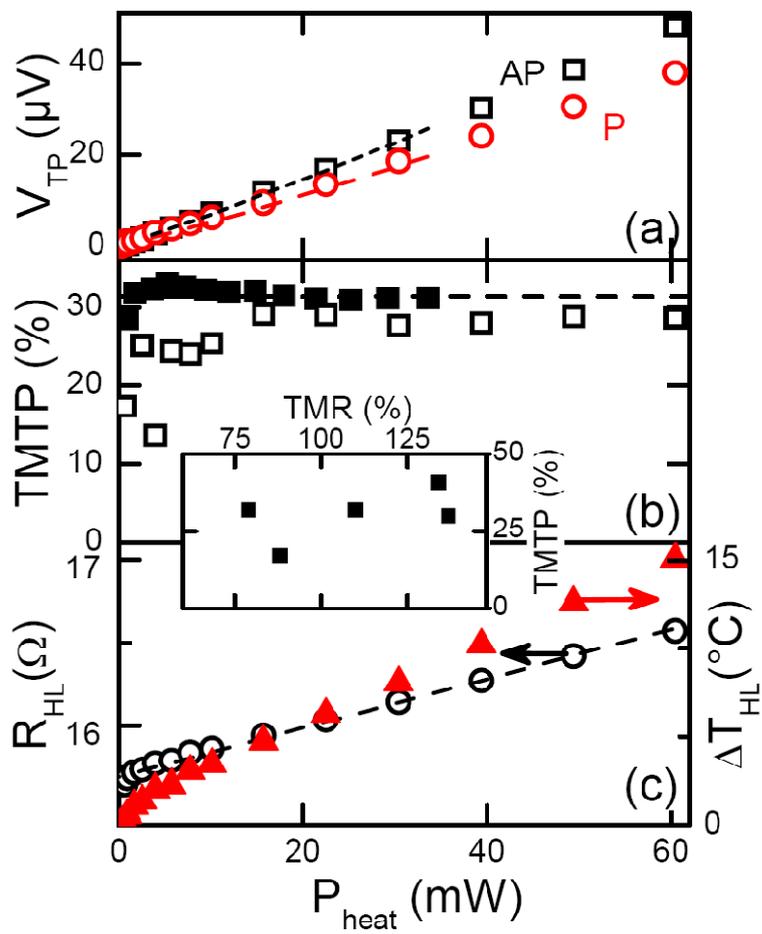

**Figure 3**

**N. Liebing et al.**

**FIGURE AND TABLE CAPTIONS:**

**Table 1:** Thermoelectric and magnetic properties of the measured MTJs: total anisotropy $\mu_0 H_K$, TMR ratio, and TMTP.

**Fig. 1:** (a) Sketch of magneto thermoelectric voltage measurements. A current $I_{heat}$ is applied through the heater line (HL) giving rise to a temperature gradient across the MTJ. The thermally induced voltage $V_{TP}$ across the MTJ nanopillar is measured. In-plane magnetic fields $H_S$ can be applied by external coils. (b) Electron micrograph of a typical device with heater line (HL, between full lines), bottom contact (BC, between dashed lines) and top contact (TC). The position of the MTJ nanopillar on the BC is indicated. (c) Simulated temperature profile in the center of the nanopillar above the BC as function of MTJ stack thickness $t$. $P_{heat}$ = 60 mW. $t$ = 0 nm corresponds to top of Cu BC. Position of MTJ is indicated.

**Fig. 2:** Magnetic field dependence of TMR and TMTP of two typical devices MTJ-3 (a,b) and MTJ-2 (c,d). (a) Easy axis TMR loop without (black) and with (red) an applied heater current of $I_{heat}$ = 38 mA. The TMR switching fields $H_C^-$, $H_C^+$ are shifted due to the HL field $H_{heat}$ by $\Delta H_C^-$, $\Delta H_C^+$. (b) Easy axis TMTP loops under applied DC heating powers of $P_{heat}$ = 21.5 mW (blue), 37.8 mW (red) and 58.05 mW (black). $H_{heat}$ is compensated to allow direct comparison to TMR loops. Black dashed line is $V_{TP,short}$ of shorted MTJ. (c) Angular dependent TMR loop for 360° in-plane rotation of free layer magnetization. Rotating static field is $\mu_0 H_S$ = 30 mT. (d) $V_{TP}$ under same field rotation as in (c). $V_{TP}$ is derived under AC excitation with $I_{heat}$ = 60 mA, $P_{heat}$ = 35.15 mW and $\pm H_{heat} \approx \pm 6$ mT.

**Fig. 3:** (a) $V_{TP}$ as function of $P_{heat}$ for parallel (P, red) and antiparallel (AP, black) orientation of MTJ-1. DC (symbols) and AC (lines) data agree. (b) TMTP ratio *vs.* $P_{heat}$ for AC (full squares) and DC measurements (open squares). Inset: TMTP *vs*. TMR of the devices of Table 1. (c) HL resistance $R_{HL}$ (left scale, circles) and increase of HL temperature $\Delta T_{HL}$ (right scale, triangles) as function of $P_{heat}$.